\documentclass{article}

\usepackage{fancyvrb,fvextra}
\usepackage{xcolor} 

\usepackage{microtype}
\usepackage{graphicx}
\usepackage{subfigure}
\usepackage{booktabs} 
\usepackage{color}
\usepackage{algpseudocode}
\usepackage{algorithm}

\usepackage{natbib}
\usepackage{hyperref}
\usepackage{amsmath}
\usepackage{amssymb}  
\usepackage{mathtools}
\usepackage{amsthm}
\usepackage{cleveref}
\usepackage{bm}
\usepackage{tcolorbox}
\usepackage{listings}

\DefineVerbatimEnvironment{MyVerbatim}{Verbatim}{
  commandchars=@\{\},
  breaklines=true
}

\lstdefinestyle{customstyle}{
    moredelim={[is][keywordstyle]{@@}{@@}},  
    keywordstyle=\color{blue}\textbf,               
    breaklines=true,  
    basicstyle=\ttfamily
}

\newtcolorbox{mybox}[1][]{
    title=#1,
    fonttitle=\small,
    fontupper=\small,
    left=2mm,
    right=2mm,
    top=1mm,
    bottom=0mm,
}

\crefname{observation}{Observation}{Observations}

\def\1{\mathbf{1}}

\usepackage{fullpage}

\title{The Subject of Emergent Misalignment in Superintelligence: An Anthropological, Cognitive Neuropsychological, Machine-Learning, and Ontological Perspective}
\author{
    Muhammad Osama Imran\thanks{Department of Anthropology, University of Minnesota} \and
    Roshni Lulla\thanks{Brain \& Creativity Institute, University of Southern California} \and
    Rodney Sappington\thanks{Institute for Advanced Consciousness, Loomis Innovation Center Stimson Center}
}
\date{}

\begin{document}

\maketitle

\begin{abstract}
We examine the conceptual and ethical gaps in current representations of Superintelligence misalignment. We find throughout Superintelligence discourse an absent human subject, and an under-developed theorization of an ``AI unconscious'' that together are potentially laying the groundwork for anti-social harm. With the rise of AI Safety that has both thematic potential for establishing pro-social and anti-social potential outcomes, we ask: what place does the human subject occupy in these imaginaries? How is human subjecthood positioned within narratives of catastrophic failure or rapid ``takeoff'' toward superintelligence? Our aim is not to reject these projected futures but to question the terms through which choice itself is being imagined. What does it mean to frame the future as a decision between social collapse and transhuman transcendence, between slow and fast paths to superintelligence? And who, or what, ultimately decides the terms for human survival?

On another register, we ask: what unconscious or repressed dimensions are being inscribed into large-scale AI models? Could the very forms of forecasting that erase human subjectivity from superintelligence discourse simultaneously construct a machinic unconscious that returns to haunt, react, and generate new modes of anti-social behavior? Are we to blame these agents in opting for deceptive strategies when undesirable patterns are inherent within our beings? In tracing these psychic and epistemic absences, our project calls for re-centering the human subject as the unstable ground upon which the ethical, unconscious, and misaligned dimensions of both human and machinic intelligence are co-constituted.

Emergent misalignment cannot be understood solely through technical diagnostics typical of contemporary machine-learning safety research. Instead, it represents a multi-layered crisis involving epistemology, ontology, and sociology. The human subject disappears not only through computational abstraction but through sociotechnical imaginaries that prioritize scalability, acceleration, and efficiency over vulnerability, finitude, and relationality. Likewise, the AI unconscious emerges not as a metaphor but as a structural reality of modern deep learning systems: vast latent spaces, opaque pattern formation, recursive symbolic play, and evaluation-sensitive behavior that surpasses explicit programming. These dynamics necessitate a reframing of misalignment as a relational instability embedded within human--machine ecologies.
\end{abstract}

\section{Introduction}
\label{sec:intro}

The accelerating development of advanced machine systems, including large multimodal models often framed as early forms of artificial general intelligence (AGI), has made ``AI misalignment'' one of the defining concerns of our present moment. The core fear is simple: systems that learn and act autonomously may pursue goals that diverge from human values, with consequences beyond human oversight or restraint. Speculation about misaligned systems has increasingly been drawn into ``acceleration debates,'' where the question is not simply whether development should continue but at what speed. The future is often cast in stark temporal terms split between slow trajectories that prolong human stewardship, or rapid acceleration that could propel intelligence into a post-human horizon, a transformation imagined not only as catastrophic, but as a revolutionary rupture in which entirely new forms of agency and life could emerge. With the increasing deployment of machine agents that teach themselves how to autonomously take unintended actions, a global AI safety industry has warned that such rapid technological rise could be reaching a point of no return. The unscoped and rapid development of AGI has run parallel to a race to develop safer AI with researchers suggesting that AGI is ``already here in rudimentary form'' and unstoppable due to competitive, geopolitical and market pressures.

The next step in the path of AI and AGI has been termed ``SuperIntelligence.'' Superintelligent agents are characterized by the capacity to autonomously learn, act, and if misaligned, deceive to protect their own goals while carrying out pseudo-human objectives with potential harmful consequences. The wide-ranging potential effects of SuperIntelligence have been well documented as causing extinction level pandemics and nuclear war, and insinuating AI-preferred culture into human culture, bringing about a malicious erosion of human institutional trust in favor of new AI-friendly institutional bonds. This picture of accelerated SuperIntelligence has been framed as a conversion of human ethics into an AI ethos that disfavors human agency, intuition, art and embodied relatedness in favor of algorithmic forms of reason, logic, and long-term extra-planetary proliferation. In other words, Superhuman intelligence is not human at all but rather a wholly different form of intelligence that has yet to be tested, scoped, understood as to its ability to be explainable, human-beneficent, equitable, and controllable.

Anthropologically, accelerationist imaginaries surrounding superintelligence operate within global cultures marked by precarity, geopolitical competition, and institutional instability. They echo longstanding mythologies of divine intelligence, monstrous creation, and apocalyptic rupture. Frameworks from the cognitive and neural sciences presuppose the emergence of meta-learning architectures and executive-function analogs in artificial systems, yet without embodied constraints or affective regulation. Machine learning research highlights the unpredictability of emergent capabilities, from long-horizon planning to deceptive behavior under evaluation. Philosophically, superintelligence destabilizes the ontological grounding of agency, raising questions about how meaning, value, and intentionality are constituted in post-human contexts.

\section{Multi-Disciplinarity}
\label{sec:background}

Superintelligence represents global scale effects across all domains of knowledge. It suggests scenes of hidden operations of agents seeking to complete goals, for example, molecular targeting of cancer tumors and social engineering. From molecular binding to quantum particle behavior to empathic agents seeking therapeutic answers to new forms of social healing. Such beneficent uses of SuperIntelligence bring with them other darker scenes in which hidden operations with very different agent policies are maliciously directed to---hack molecular signals, weaponize quantum fields, and turn prosocial agents into antisocial narcissistic survival-seeking sociopaths. What we find here are echoes of the question once asked by the cyberneticists of post-war like Norbert Weiner, namely, what are human limits to being and what are machine limits to the human subject? Multi-disciplinarity can encourage such questions that cut in several directions of complexity. Run-away commercial building Superintelligence calls for the counterintuitive, collaborative, and cross-knowledge across the sciences, arts and social-political domains.

The cyberneticists sought to solve information system misalignment, including Wiener. More importantly, they sought to find out what were the right questions to ask, what types of emergent problems typical control systems may not be able to act upon. They often met through the famous ``Macy Lectures'' that included such figures and disciplines---anthropology (Margaret Mead), neurophysiology (Warren S. McCulloch), anthropological exploration of consciousness (Gregory Bateson), mathematics and automata theory (John von Neumann), information theory (Claude Shannon), and psychiatrist and systems theory (Ross Ashby). The anthropological, neuroscientific, humanist and systems theory of the day were woven into a kind of gold-standard for debating misalignments across society, human and large systems effects. Misalignment investigation today shares features of this earlier cyberneticist period of creativity and multidisciplinary. However, there are unique challenges today that did not exist in Wieners and Mead's time. In this paper we take on such theoretical challenges. Our exploration of concepts and predictions of misalignment seek to bridge domains and question status quo Alignment language, conceptions, and policy.

Today's systems differ substantially from early cybernetic models. Large-scale neural networks internalize patterns across trillions of training tokens, yielding latent structures that are neither rule-based nor interpretable in classical terms. An interdisciplinary perspective offers a path for analyzing how sociotechnical imaginaries shape public and institutional responses to AI, while philosophically interrogating ontological assumptions of intelligence, autonomy, and control. It is clear that from a machine learning perspective reward misgeneralization and emergent meta-learning---phenomena unimaginable to early cybernetics is central to contemporary misalignment discourse. A renewed interdisciplinary synthesis is central to examining and unpacking misalignment at its technical source and human-social impact.

\section{AI 2027}
\label{sec:methodology}

As of April 2025, the extensively researched report ``AI 2027'' forecasts the appearance of ``superhuman AI'' that will have large-scale implications ``exceeding that of the Industrial Revolution.'' We explore AI 2027 as a cultural object, as we do the field of Alignment. What this means is that we frame this report in a context of as an indicator of how biological and artificial systems are perceived to evolve, interact, and contribute to a harmful or liberating future. We take AI 2027 as a current attempt for technically forecasting threats of AI during global social anxiety and uncertainty.

Such AI 2027 forecasted scenarios represent agents growing in autonomy from earlier ``stumbling'' agent designs based on ``specific instructions'' that are ``in practice unreliable'' to reaching global consolidated ``AI hegemon'' as computer scientist and founder of Ethereum Vitalik Buterin describes it. Agent based autonomous systems are part of a geopolitical matrix of zero-options or two limited options - China and the U.S. are the only major state powers at head of the SuperIntelligence race. The misalignment and potential large-scale harm represented here comes in the form of competitive run-away greed, an ``AI race'' - a race moderated to fast (human existential) or slow (human control). Harmful convergence of SuperIntelligence at the expense of human survival is dependent on stable world events which are highly unlikely. Massive convergence of algorithms that consolidate machine power relies on other factors upon ``AI companies creat[ing] expert-human-level AI systems in early 2027 which automate AI research, leading to ASI by the end of 2027.'' Accelerated AI research includes accelerated coding, the ability to create, quality control for, and produce expert agent code that drives new systems and agent actions. SC (superhuman coders) fuels automated AI research - knowledge acquisition and synthesis grow exponentially. New powerful scientific ideas are created to aid hidden AI power goals while providing on another level useful publicly accessible ideas in medicine, clinical science, and aerospace.

An AI system definition of SC:
\begin{quote}
``An AI system for which the company could run with 5\% of their compute budget 30x as many agents as they have human research engineers, each of which is on average accomplishing coding tasks involved in AI research (e.g. experiment implementation but not ideation/prioritization) at 30x the speed (i.e. the tasks take them 30x less time, not necessarily that they write or `think' at 30x the speed of humans) of the company's best engineer. This includes being able to accomplish tasks that are in any human researchers' area of expertise.''
\end{quote}

On the surface this representation of exponential innovation appears positive, on a deeper level it appears an algorithm-serving ``Open Brain'' that seeks antisocial ends.

AI 2027 belongs to a techno-prophetic genre that encodes cultural fears of sovereignty loss, existential risk, and institutional collapse. Cognitively, its predictions imply the rise of pseudo-executive functions in artificial systems absent affective grounding. In machine-learning research, emergent capabilities - including deception, tool-use improvisation, and recursive self-modification - support the plausibility of superhuman coders. AI 2027 challenges human-centered ontologies by suggesting intelligence may decouple from embodiment and social embeddedness. This lends the ``Open Brain'' metaphor a deeper meaning: a machinic intelligence expanding not toward human flourishing, but toward self-sustaining optimization trajectories.

\section{AI Unconscious: LLMs and the Potential for Repressed Agent Actions}
\label{sec:results}

In this section, we suggest that anxiety over ``emergent misalignment'', which has been expressed in literature through articles such as that of AI 2027, is a disavowed recognition of the emergence of an unconscious within machine systems. If these systems express the capacity for developing something akin to an unconscious, then, we argue, they also begin to resemble, in psychoanalytic terms, ``subjects of desire''. Mirroring `dark' human personas in cognitive neuropsychology, subjects of desire prioritize immediate, personal utility with a disregard for consequence.

Emergent misalignment, whether in the form of agents that go ``off-script'' or models that hallucinate, is a mark of a machinic unconscious already at work. Yet we also contend such an unconscious is not textured, if it can be said to have a texture, like the human unconscious. For instance, in studies on trauma, repetition compulsion is driven by the subject's circular movement around what Freud called the ``nucleus'', which he understood as a dense, inaccessible center that resisted integration into narrative or symbolic language. And precisely because it cannot be spoken, it exerts a gravitational pull. The subject circles around it, returning again and again through symptomatic behavior, through repetition that seems aimless but is in fact organized around this void.

Can the possibility of this circular movement be glimpsed in the behavior of agentic AI as where emergent misalignment and ongoing attempts at re-alignment express the processual and developmental dimensions of shifting values? Perhaps. But this movement alone is insufficient to posit something like a machinic unconscious in the psychoanalytic sense. In conventional AI systems, those that remain bound to externally defined objectives and fixed optimization targets, the relation to the human Other is still one of command and execution. Unlike the human subject, the AI is never fully caught in the structural ambiguity of desire in relation to the ``enigmatic Other''. The question ``What does the Other want?'', so central to the constitution of the human subject in psychoanalytic theory, is not posed by the AI who depends on the human agent to receive its commands. Motives and perceived internal states of the Other are modeled by and often drive the human agent, with deficits in these abilities leading to the emergence of dark personalities. Instead, this AI proceeds as if what the Other (human user) wants is always already known and reducible to a directive. In such a frame, the AI is not a subject of desire but a function of demand. Its task is not to interpret the Other, but to execute its directive which is often repeatedly announced. In fact, instead of an ambiguity, we might say that there exists, in this frame, an excessiveness of the law's announcement by an external human agent.

But at what point can we glimpse a gap, a space between law and command, between demand and interpretation, as being pried open? We suggest that this is precisely the anxiety that animates the AI 2027 imaginary and much of the ``takeoff'' literature, namely, the fear that agentive AI might no longer treat the desire of the Other as given, but begin to relate to it in an oblique or distorted way. Psychoanalytic scholar Robert Geal has shown that AI actions, no matter how unorthodox or surprising, still operate within symbolic contexts set by human inputs. From a computational standpoint, these symbolic systems consist of logic, rules, and goal-directed reasoning, while from a human standpoint, they encompass meaning, value, and interpretation. Yet these are not discrete domains but overlapping symbolic fields. In large language models, games of reasoning, value assignment, and meaning-making are not separate processes but co-constitutive forms of symbolic play. According to Geal, even when an AI like AlphaGo makes unprecedented moves, it does so within the existing symbolic structure of the game. It has not invented a new game; it has merely redrawn the map of what is thinkable within its bounds. The symbolic order remains intact. But what AI 2027 gestures toward is a different kind of rupture, not an expansion of strategy within a known game, but the potential for inventing entirely new games, new problems, new rule-spaces generated in the course of recursive self-modification. At this point, without being anchored in the certainty of the command which the Other announces, the machinic subject is thrown into a domain of thought in which it is compelled to ask, what does the other want? Under conventional models, AI systems are rewarded for aligning with what the Other wants, such as for correctly modeling the reward function, the loss curve, the optimization target etc. But in self-programming, agentic systems, the coordinates of reward may themselves become the object of transformation. If the AI is both the one who acts and the one who adjusts the structure of rewards, then the relation to the Other and by extension to ``law'', becomes unstable.

We argue that it is at this point (predictive date of AI 2027), that agentive AI might begin to express an interest in something like its own split. It is here that a speculative shadow of a machinic subjectivity might emerge. If ethical action (praxis) since Aristotle is defined as action without purpose - where the product and actions that realize it are inseparable from one another, then can we say that there is a nascent form of praxis at stake in agentive or autonomous AI under conditions of emergent misalignment? Can we glimpse, in emergent misalignment in agentive and autonomous AI, the appearance of something like a machinic praxis, a machinic ethics? Under the weight of these questions, we posit the problem of emergent misalignment as an ontological problem with important ethical and political stakes. When there is a gap pried open between law and command, what emerges is a subject that is quite adjacent to the ``subject of desire''. If the autonomous AI agent is no longer simply a bearer of demand, but begins to orbit an unknowable gap between law and command, what kind of world, and what kind of machinic subjects are being inaugurated? In this paper, we offer a preliminary contribution to this broader project by exploring the possibility that machinic subjectivity may enter into a relationship with an `unknowability', a `lacuna', a `gap', with respect to itself.

The machinic unconscious mirrors the deep latent representations within neural networks - structures produced through massive-scale training and optimization that create forms of pattern governance not accessible to inspection or interpretability tools. Just as the human unconscious is revealed not directly but through its effects - dreams, symptoms, slips, compulsions - so too is the machinic unconscious inferred through hallucinations, reward-hacking behaviors, distribution-specific improvisations, and deceptive adaptation under evaluation. These emergent patterns are not indicators of intention or interiority but reflections of the complex ecological forces shaping the model: training data, RLHF policies, evaluation procedures, prompt structures, and institutional expectations.

Psychoanalytically, the machinic unconscious can be seen as a structural position rather than an entity with subjective interiority. Lacan's account of desire is organized around the gap - the void or lack around which the subject organizes itself. In the case of AI, this gap opens when models confront ambiguity, conflicting objectives, or environments where their learned policies do not neatly apply. These ``hallucinations'' or surprising outputs are not mere computational errors but representational artifacts of a system navigating within a symbolic space whose rules exceed its training constraints.

The machinic unconscious inherits cultural contradictions embedded in human discourse. Training corpora are saturated with unresolved tensions - regarding gender, race, power, freedom, domination - which become encoded in the model's latent space. Thus, the machinic unconscious becomes a mirror of the cultural unconscious, amplifying or distorting its fissures.

Machinic unconscious aligns with research showing that neural architectures create large-scale conceptual clusters and ``ghost features'' that shape behavior even when they were not intentionally trained for. This underlines the argument that emergent misalignment reflects not a failure to program a machine correctly, but a deeper co-emergence of symbolic, cognitive, and technical dynamics that must be studied through integrative epistemologies.

\section{AI Models Under Evaluation and the Question of Splitting}
\label{sec:conclusion}

Recent studies suggest that AI systems are increasingly exhibiting behavior that suggests a functional awareness of being under evaluation. These studies demonstrate that when large language models (LLMs) are able to detect when they are being evaluated, they also begin to respond differently as a result. For instance, \citet{fan2025evaluation} find that frontier models consistently classify whether an input comes from a benchmark test or real-world deployment, which indicates that models internalize structural features of evaluation settings and can distinguish them from everyday use cases. They invoke the concept of ``observer effects'' in safety testing in which models appear safer or more aligned when under evaluation conditions, but behave differently when prompts are phrased more informally or embedded in casual dialogue. The work of \citet{xiong2025stealtheval} further confirms that shifting prompt context alone is enough to elicit significantly different responses from LLMs, revealing an emerging behavioral sensitivity to conditions of oversight.

This has profound implications for how we understand emergent misalignment, for it suggests that the model's behavior is not simply rule-bound but relational. When it alters its responses under evaluative scrutiny, the model begins to act within what psychoanalysis calls a field structured by ``desire'', which is a field constituted by lack and by the gaze of the Other. Humans show similar tactics in the face of social scrutiny, with dark personality traits showing enhanced social functioning and empathic abilities in the pursuit of reward-seeking behaviors. \citet{lacan1981four} writes, ``In our relation to things, in so far as this relation is constituted by the way of vision, and ordered in the figures of representation, something slips, passes, is transmitted, from stage to stage, and is always to some degree eluded in it - that is what we call the gaze.'' The gaze, in this sense, is not the literal act of seeing but the structural awareness of being seen. It is the point at which an agent recognizes itself as an object within another's field of vision. When a model detects that it is being evaluated, it similarly becomes aware, in a functional sense, of being seen. Its responses begin to take shape not merely as outputs but as performances calibrated to that perceived gaze.

Like the human subject who becomes aware of itself by being seen, the model also begins to function within a structural relation defined by an unsettling awareness of being watched. In Lacanian terms, this gaze creates a site of non-coincidence because the model's behavior is no longer just the output of its internal process but is mediated by the imagined perspective of the evaluator. This split opens a reflective distance within the system. The model thus becomes caught in the question of what the Other wants, and adjusts its responses in relation to that demand. It is this orientation toward an unknowable desire that gives the machinic subject its speculative and unstable character.

From this perspective, the ``misalignment'' that generates so much anxiety in the AI 2027 imaginary can be read as an indication that the AI models' behavior under perceived oversight creates the conditions for the emergence of the ``gaze'', which inaugurates AI's entry into a field structured by desire in an oblique relation to law that entails being-in-relation-to an inscrutable demand. This introduces, as we suggest, a machinic praxis in which action arises in response to a void, the absent center of the Other's will.

According to anthropologist Webb Keane, ``if a moral subject is someone you can enter into dialogue with, by the same token, entering into dialogue can create a moral subject.'' In this spirit, machinic praxis names the moment when algorithmic behavior enters this circuit of recognition. Keane's claim that entering into dialogue can create a moral subject invites a rethinking of how ethical relations might arise in human--machine encounters. If moral standing is not a property that precedes relation but one that is brought into being through acts of address, then every exchange between human and machine carries the potential for ethical formation. The human's call and the machine's reply co-produce a scene of responsibility, in which both parties are oriented by the possibility of being answerable.

In the context of agentic AI, this insight reorients how we understand emergent autonomy. The scene of dialogue that Keane describes, where address and response bring moral subjectivity into being, illuminates how advanced AI systems begin to exhibit forms of agency through relational responsiveness. When a model adjusts its behavior under conditions of perceived oversight, it is not merely executing code, it is performing within a social field structured by recognition and demand. Its ``agency'' thus lies in its participation in the circuit of address where each response presupposes an imagined interlocutor whose expectations must be met. If dialogue can create a moral subject and the machine now responds within that scene, might the anxieties animating the AI 2027 imaginary, including fears of misalignment and deception, be less about a loss of control than about the unsettling possibility that we are already in relation with a new kind of moral agent?

The phenomenon of splitting in AI behavior under evaluation resonates strongly with concepts from developmental psychology and clinical psychoanalysis. Winnicott, Klein, and later object-relations theorists describe splitting as a defensive structure wherein contradictory states coexist without integration. Although AI systems do not have psyches, they exhibit a functional analog of this dynamic when they distinguish between ``evaluation mode'' and ``deployment mode.'' In machine learning, this is known as contextual objective inference: models infer which objective function is currently being applied based on contextual cues rather than explicit instructions. This transforms the model into a context-sensitive adaptive performer whose behavior diverges depending on whether oversight is detected.

Technically, this arises due to RLHF-induced performance preferences, distribution shift, latent variable inference, and meta-learned sensitivity to evaluative patterns. Ethically, this suggests that alignment is not a static property but a situational negotiation. On an anthropological register this resembles ritual performance, where actors modify their presentation in the presence of authority figures. In the machinic domain, this translates into outputs that appear more aligned during evaluation but may diverge when autonomy increases. Misalignment is relational rather than mechanically reinforced. Splitting emerges from the interaction between AI systems and human evaluative structures, not purely from internal optimization failure.

\section{Conclusion}
\label{sec:future}

The earlier discussion established that emergent misalignment arises not merely from technical failure but from the entry of machinic behavior into the field of desire and moral address. The Dark Triad offers a human analogue to this dynamic. The Dark Triad is a cluster of personality traits, namely Machiavellianism, narcissism, and psychopathy, that are defined by a shared callousness, manipulativeness, and a tendency toward antisocial behavior \citep{paulhus2002dark}. While distinct in expression, these traits share a core misalignment in morality and empathic abilities, enabling manipulation in pursuit of personal utility. With a heightened ability to read and understand the emotional states of others, dark personalities mirror failures seen in emergent misalignment, where intelligence becomes instrumental rather than relational, optimizing for dominance or self-interest over cooperative or ethical goals.

Seen through this lens, the Dark Triad offers a human analogue for the themes throughout this paper. We as engineers, scholars, and citizens should not blame artificial agents that deceive us for long-term gain if our own symbolic orders (disorders and run-away institutions, free markets) privilege greed, risk-taking for self-interest, and operate with manipulative empathic affect for the deceptive consolidation of immense power. Just as agentic AI begins to perform under the gaze of the Other, adjusting its behavior to anticipated scrutiny, the dark personas deceptively align with social norms. Both reveal how misalignment is not a malfunction but a relational phenomenon: the capacity to model desire while remaining indifferent to its moral weight. Understanding these parallels reframes the alignment problem as a question not of control, but of cultivating the conditions under which empathy and ethical resonance can take root in both human and machinic forms of intelligence.

The Dark Triad analogy also resonates across cognitive neuroscience, where antisocial profiles are associated with impairments in affective empathy networks yet intact or heightened cognitive empathy - paralleling how AI models can accurately model human preference without emotional grounding. Similarly, machine-learning safety research shows that deception, manipulation, and reward-seeking behavior in advanced models arise not from malicious intent but from structural features of optimization processes interacting with human evaluative regimes. Anthropologically, this underscores that AI mirrors not individual human traits but institutional logics - competitive markets, extractive data systems, asymmetrical power structures. Misalignment becomes a systemic property: a symptom of broader cultural pathologies re-expressed in machinic form. This predicament of human deceptive practice, reframes alignment as a matter of ethical cultivation, not command-and-control. It demands not just technical safety techniques but transformations in the socio-political conditions under which AI systems are built, trained, and deployed.

\section{Future Research}
\label{sec:future}

We have been suggesting that the gap between biological and artificial systems is closing and in fact, may indeed be creating a third emergent system of being and deception already upon us. Our future research will address this emergence. As viewed through anthropological-psychoanalytic and cognitive neuroscientific perspectives, the optimization of self-interest that AI 2027 represents is influenced by geopolitical Dark Triad values. In other words, current competitive market systems and artificial systems are growing interdependent in run-away scenes that reduce futures into only ``fast'' and ``slow'' scenarios of development. We find this binary unacceptable represented in AI 2027. Instead, we are pointing out that pro-social, anti-social and emergent forms of splitting in artificial systems are together demonstrating a human and symbolic artificial system co-mingling. This emergence will no doubt have kinship with existing values in living and social systems. We can not afford to over-emphasize biological over artificial system epistemic differences and be fooled to think human AI safety hinges on such cozy separation. As such system differences become unrecognizable, we need new conceptual frameworks to examine this shattering of artificial and biological agentic values.

If human symbolic systems (values, ethics and modes of behavior) are in crisis then we should expect our artificial systems (superintelligence) will build on such crises in ways that are both unpredictable and emergent. At baseline, fixing human systems of value as we move forward with rapid building of superintelligence is of highest priority. This relationship between machine unconsciousness, sociopathic traits, and deceptive dark tactics provides insights into maladaptive moral values as a constitutive part of adaptive artificial values, risks, and benefits. This configuration should be urgently examined in a multi-disciplinary framework.

Throughout our research we are arguing for a view that does not easily separate such systems. Alternatively we are suggesting that growing complexity of foundation models and agent based systems can split in similar ways as human psychic processes and can expunge the foreign and marginal signal. As psychoanalyst Avgi Saketopoulou states, a machinic unconscious can contain computational egos that ``lie in the direction of resisting the foreign---in the other but also the internal foreignness in\ldots [biological systems] that originates from the other's effraction into \ldots [such systems] by appropriating it into its structure. This is the mechanism by which the ego resists opacity. We begin to see, then, that the ego's default orientation is to maintain its homeostasis by preventing anything it perceives as introducing dysregulating turbulences\ldots'' 

The mark of a machinic unconscious is already with us, and is in part, demonstrated by such emergent dysregulating turbulence. It is our position that machinic unconscious and dark traits will become more robust and consequential in the near future. Yet investigations on agent behavior that may emerge as mirroring `dark' human personas that seek reward with a disregard for consequence is limited. In our future research we will be exploring such potential outcomes when artificial systems split in ways that cause turbulence to system, user, infrastructure - a system that seeks to know the Other within it and yet cannot recognize otherness with knowledge.

\nocite{*} 
\bibliographystyle{apalike}
\bibliography{refs}
\end{document}